\documentclass[aps,prl,twocolumn]{revtex4}
\usepackage{graphicx}

\begin{document}

\title{Low-field octupoles and high-field quadrupoles in URu$_2$Si$_2$}
\author{Annam\'aria Kiss}
\author{Patrik Fazekas}
\affiliation{
Research Institute for Solid State Physics and Optics,
Budapest 114, P.O.B. 49, H-1525 Hungary.}
\date{\today}

\begin{abstract}
The recent experimental finding of large-amplitude antiferromagnetism induced
by uniaxial strain shows that the "hidden" low-field order of 
URu$_2$Si$_2$ breaks time reversal invariance. We propose a new crystal 
field model which supports $T_z^{\beta}$ octupolar order in the low-field
phase, and quadrupolar order in a disjoint high-field phase. The 
temperature dependence of the linear and third order magnetic 
susceptibility is in good agreement with the observed behavior. 
 \end{abstract}
\maketitle

The nature of the $T_O=17.5{\rm K}$ phase transition of URu$_2$Si$_2$
is a long-standing puzzle \cite{harrison}. Though URu$_2$Si$_2$ was long 
considered as a "light" heavy fermion system, implying that the $f$-states 
should be included in the Fermi volume, many aspects of the normal state 
behavior are well described in terms of a
localized $f$-electron model. 
Specific heat measurements \cite{schlabitz} show
that an electronic entropy of $O(\ln{2})$ is released by the time the
temperature reaches 30K, and a sizeable fraction of it is associated with the 
$\lambda$-anomaly at 17.5K. Thus the phase transition should be 
associated with the full-scale ordering of a localized degree of 
freedom per site, 
but the nature of the order parameter remains hidden. It is obviously not the 
tiny ($M_z\sim 0.03{\mu}_{\rm B}$, $z$ being the tetragonal fourfold axis 
[001]) 
antiferromagnetic moment  observed by
neutron scattering \cite{broholm}. In fact, the association of  micromagnetism 
with the 17.5K transition is  dubious, since it depends on sample quality, while 
the thermodynamic transition itself is a robust sharp feature. 

{\sl If} one assumes that the observations are made on single-phase specimens, 
then the weak antiferromagnetism should be described as a secondary phenomenon 
driven by the primary ordering of the hidden order parameter. The staggered  
dipole moment and the hidden order parameter would possess the same spatial 
and time reversal symmetry.  The finding of a first order transition to 
a high-pressure phase with large moments was argued to favour this scenario 
\cite{flouq}. 
A general symmetry analysis listed local octupoles as well as 
 triple-spin correlators \cite{agterberg}. We note, however, that an extensive mean 
 field study by Santini et al considered the possibility of octupolar order, and discarded it in favor 
 of the alternative of quadrupolar order \cite{Santini}. Most recently, 
unconventional 
 density waves with alternating plaquette currents were advocated \cite{dw}.
 The plaquette current can give rise to weak orbital antiferromagnetism, and is thus in 
 principle well suited to describe URu$_2$Si$_2$. However, these works are based on the 
 extended $s$-band Hubbard model; we prefer a description emphasizing the orbital character 
 of $f$-electrons.

We propose an alternative scenario in which hidden order and 
antiferromagnetism  are of  
 different symmetries. There is experimental evidence that the
apparently tiny moments belong to a minority phase, and the hidden order of
the majority phase is  non-magnetic \cite{matsuda,yokoyama}. 

Assuming that hidden order is {\sl not} of the same symmetry as $J^z$ 
dipoles, there are still two 
basic options: hidden order may, or may not, break time reversal 
invariance \cite{shah}. 
Earlier, both possibilities seemed open \cite{bernal}, 
and Santini's quadrupolar model gives 
an example of time reversal invariant hidden order.  However, we are going to
argue that recent observations necessitate to postulate  
time reversal invariance breaking hidden order.

According to a crucial recent experiment \cite{yokoyama}, uniaxial pressure 
applied in  [100] or [110] directions induces a relatively 
large magnetic moment in direction [001].  In contrast, 
stress in the [001] direction does not induce any sizeable moment. Since 
stress is a time reversal invariant perturbation, it can induce magnetism only
from an underlying (hidden) order which itself breaks time reversal
invariance. The directionality of the effect indicates time reversal
invariance breaking orbital order, most straightforwardly octupolar order.

Octupolar order as  primary order parameter
was suggested for Phase IV of Ce$_{1-x}$La$_x$B$_6$ \cite{KK}, and
for NpO$_2$ \cite{npo2}. 
Knowing such precedents, it is a plausible idea to check 
whether a difficult-to-identify order of an $f$-electron system 
is  octupolar. 

The tetragonal symmetry classification of the local order parameters 
\cite{shiina97}  in the 
absence of a magnetic field is given in Table~\ref{tab:tab1}.  $g$ and $u$ refer to "even" and 
"odd" under time reversal. Here we neglected wave vector dependence; 
it will be specified when we mean staggered rather than uniform moments. 

\begin{table}[ht]\caption{Symmetry classification of the local order parameters for 
${\bf H} =0$ (${\cal D}_{4h}$ notations\protect \cite{inui}, overline means
symmetrization \cite{shiina97}).}
\label{tab:tab1}
\centering
\begin{tabular}{|c|c||c|c|}
\hline \hline sym ($g$)& operator & sym ($u$) & operator\\[1mm]
\hline 
$A_{1g} $ &  ${\cal E}$ & $A_{1u} $&$ {\overline{J_x J_y J_z (J_x^2 - J_y^2)}} $\\[1mm]
$A_{2g}$ & ${\overline{J_x J_y  (J_x^2 - J_y^2)}} $& $A_{2u}$ & $J_z$ \\
$B_{1g} $ & ${\cal O}_2^2$ &  $B_{1u} $ &  ${\cal T}_{xyz}= {\overline{J_x J_y J_z }}$\\
$B_{2g}$ & ${\cal O}_{xy}={\overline{J_x J_y }}$ & $B_{2u}$ &  ${\cal T}^{\beta}_z={\overline{J_z(J_x^2-J_y^2)}}$\\
$E_g$ & $\{ {\cal O}_{xz}, {\cal O}_{yz} \}$ & $E_u$ & $\{ J_x, J_y \}$  \\
\hline
\end{tabular}
\end{table}

We argued that the 
$H=0$ order parameter must be one of the $u$ operators; 
it cannot be $A_{2u}$ or 
$E_u$ because that would mean magnetic order; later we
mention why it cannot be $A_{1u}$; so 
it must be one of the octupoles ${\cal T}^{\beta}_z$ ($B_{2u}$), 
or ${\cal T}_{xyz}$ ( $B_{1u} $). 
Lacking a microscopic analysis of the multipolar interactions in 
URu$_2$Si$_2$, we cannot decide between the two, and arbitrarily choose 
the ${\cal T}^{\beta}_z$ octupole as {\sl the} zero-field order parameter 
\cite{xyz}.

Switching on a field ${\bf H} \parallel {\hat z}$, geometrical symmetry is lowered to ${\cal C}_{4h}$. 
However, the relevant symmetry is not purely geometrical. Though taken in itself, reflection in the 
$xz$ plane $\sigma_{v,x}$ is not a symmetry operation 
(it changes the sign of the field), 
combining it with time reversal ${\hat T}$ gives  the symmetry operation 
${\hat{\boldmath T}}\sigma_{v,x}$.  The same holds 
for all vertical mirror planes, and ${\cal C}_2\perp {\hat z}$ axes, 
thus the full symmetry group 
consists of eight unitary and eight non-unitary symmetry operations 
\cite{inui}
\begin{equation}
{\cal G} = {\cal C}_{4h} + {\hat{\boldmath T}}\sigma_{v,x} {\cal C}_{4h} \, .
\end{equation}
We may resort to a simpler description observing that 
\begin{equation}
{\tilde{\cal G}} = {\cal C}_{4} + {\hat{\boldmath T}}\sigma_{v,x} {\cal C}_{4}
\end{equation}
is an important subgroup of ${\cal G}$, and we can base a  symmetry 
classification on it. The multiplication table of ${\tilde{\cal G}}$ 
is the same as that of ${\cal C}_{4v}$, and therefore 
the irreps can be given similar labels. It
is in this sense that the symmetry  in the presence of a field 
${\bf H} \parallel {\hat z}$ can be regarded as ${\cal C}_{4v}$ (a convention
used in \cite{shiina97}). 
The symmetry classification of the local order parameters 
valid in ${\bf H} \parallel {\hat z}$ is given in Table~\ref{tab:tab2}. The results  make it explicit that the magnetic field mixes dipoles with quadrupoles, quadrupoles with certain octupoles, etc.

\begin{table}[hb]
\caption{Symmetry classification of the lowest rank local order parameters for 
${\bf H} \parallel {\hat z}$ (notations as for ${\cal C}_{4v}$ 
\protect\cite{inui})}
\label{tab:tab2}
\centering
\begin{tabular}{|c||l|}
\hline \hline Symmetry & \ \ basis operators\\
\hline $A_1$ &  \ \ 1, $J_z$\\
$A_2$ & \ \ ${\overline{J_xJ_y(J_x^2-J_y^2)}}$, 
${\overline{J_xJ_yJ_z(J_x^2-J_y^2)}}$\\
$B_1$ & \ \ ${\cal O}_2^2$, ${\cal T}^{\beta}_z$\\
$B_2$ & \ \ ${\cal O}_{xy}$, ${\cal T}_{xyz}$ \\
$E$ & \ \ $\{J_x, J_y\}$, $\{ {\cal O}_{xz}, {\cal O}_{yz}\}$\\
\hline
\end{tabular}
\end{table}

\begin{figure}[ht]
\centering
\includegraphics[height=6cm,angle=0]{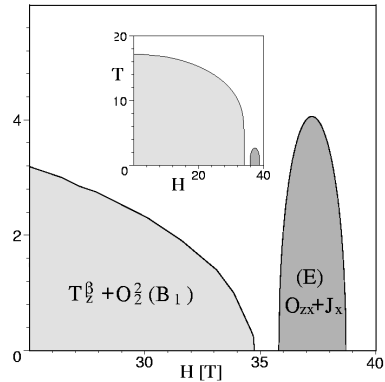} 
\caption{ The high-field part of the $T=0$ phase diagram of the multipolar model ($H$ in units of T 
(Tesla)). Vertical axis: $\langle {\cal T}^{\beta}_z\rangle$ for the low--field phase, and 
$\langle {\cal O}_{zx}\rangle$ for the high-field phase. The field-induced
mixing of the order parameters is shown within the shaded areas. 
The overall appearance of the $T$-$H$ (inset, $T$ in units of K) phase diagram is very similar. (The critical temperature of the $E$ phase is scaled up 3-fold).
\label{fig:phase}}\end{figure}

In a field ${\bf H} \parallel {\hat z}$, there can exist ordered phases with four different local symmetries: $A_2$, $B_1$, $B_2$, and $E$. The zero-field $B_{2u}$-type ${\cal T}^{\beta}_z$ octupolar order 
evolves into  the $B_{1}$-type ${\cal T}^{\beta}_z$--${\cal O}_2^2$ mixed
octupolar--quadrupolar order (Figure~\ref{fig:phase}). Experiments tell us 
that  $B_{1u}$ (and also $B_1$) order is alternating (${\bf Q}=(111)$). 
The gradual suppression of
octupolar order under field applied in a high-symmetry direction is
well-known, e.g., from \cite{kf}.  In our calculation, the octupolar
phase is suppressed at $H_{\rm cr,1} 
\approx 34.7{\rm T}$ (Figure~\ref{fig:phase}). 

There can be other kinds of order, but they cannot coexist with $B_1$ because they carry different symmetry labels. The phases can be disjoint, separated by non-ordered regimes, or when they press against each other, the transition must be first order. It is a question of detail whether isolated phases are bounded by critical lines or first order boundaries. 

Seeking agreement with high-$T$ and large-field data we postulate a
crystal-field model in which two levels tend to cross at $H>H_{\rm cr,1}$, and
they are connected by matrix elements of $E$ operators. Consequently, 
we find a high-field  $E$ phase where 
$\{J_x, J_y\}$-type transverse dipolar order is mixed with 
$\{ {\cal O}_{xz}, {\cal O}_{yz} \}$-type quadrupolar order (see
Figure~\ref{fig:phase} and Table~\ref{tab:tab2}). The overall appearance of
our phase diagram closely resembles the results of high-field measurements
\cite{jaime}.  Some experiments  identified additional domains in the $H$--$T$
plane 
\cite{kimsuslovharr}, but we think that the two phases shown in  
Figure~\ref{fig:phase} are the most robust part of the phase diagram. 

\begin{figure}[ht]
\includegraphics[height=4cm]{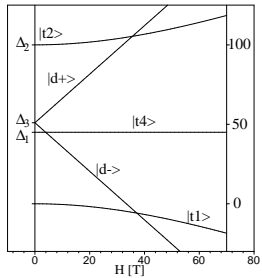}
\caption{The magnetic field dependence of the single-ion levels.\label{fig:levels}}
\end{figure}

We assume stable $5f^2$ valence, and Hund's rule $J=4$ ground state. 
Let us seek a plausible level scheme to support the postulated 
ordering phenomena. It is accepted that the ground state is a singlet, 
and it is connected to another singlet  across a gap of $\sim 100$K 
by a matrix element of $J^z$ \cite{broholm}. In our scheme, 
$\left|t_1\right>$ is the ground state, and $\left|t_2\right>$ the
$\Delta_2=100$K excitation. We need the low-lying ($\Delta_1=45$K)
singlet $\left|t_4\right>$ 
to allow induced octupole order. Finally, as in previous 
schemes \cite{Santini}, at least two more states are needed to fit 
magnetization data up to 300K. 
We found it useful to insert one of the doublets 
($\left|d_{\pm}\right>$). Level positions were adjusted to get good overall
agreement with observations but we did not attempt to fine-tune the model. 
The list of the relevant crystal field states is 
given in Table~\ref{tab:J4states} (we use $a=0.98$, $b=0.22$). 
The field dependence of the levels is shown in Fig.~\ref{fig:levels}.

\begin{table}[ht]
\caption{Tetragonal crystal field states used in the model}\label{tab:J4states}
\centering
\begin{tabular}{|l|c|c|c|}
\hline \hline state & form & symmetry & energy[K]\\
\hline $\left|t_2\right>$ & $1/\sqrt{2}(\left|4\right>-\left|-4\right>)$ & 
$A_2$ & 100\\
$\left|d_{\pm}\right>$ & $a\left|\pm 3\right>-\sqrt{1-a^2}
\left|\mp 1\right>$ & $E$ & 51\\
$\left|t_4\right>$ & $1/\sqrt{2}(\left|2\right>-\left|-2\right>)$ & 
$B_2$ & 45\\
$\left|t_1\right>$ & $b(\left|4\right>+\left|-4\right>)+\sqrt{1-2b^2}
\left|0\right>$ & $A_1$ & 0\\
\hline
\end{tabular}
\end{table}

\begin{figure}[ht]
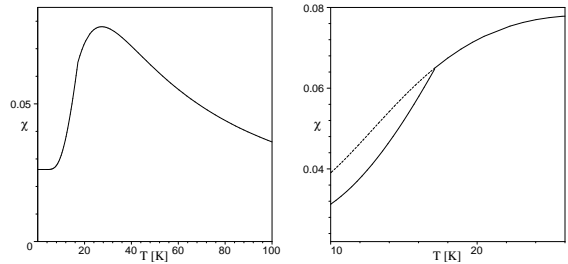

\centering
\includegraphics[height=3.8cm]{kf_urusi_fig3_left.ps}
\includegraphics[height=3.8cm]{kf_urusi_fig3_right.ps}
\caption{Linear susceptibility per site (in $\mu_{\rm B}/{\rm T}$) on extended temperature scale 
(left), and in the vicinity of the octupolar transition (right). The dashed
line gives the single-ion result. \label{fig:chi1}}
\end{figure}

\begin{figure}[ht]
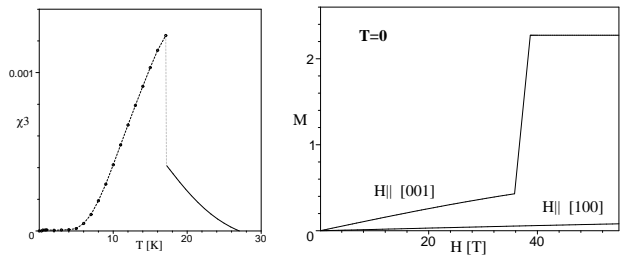

\centering
\includegraphics[height=3.6cm]{kf_urusi_fig4_left.ps}
\includegraphics[height=3.6cm]{kf_urusi_fig4_right.ps}
\caption{{\sl Left}: Nonlinear susceptibility $\chi_3$ at the octupolar
    transition. {\sl Right}: the magnetization curve at $T=0$ ($M$ in units of
    $\mu_{\rm B}$). 
\label{fig:chi3}}
\end{figure}

Our crystal field scheme differs in essential details from previous 
ones \cite{Santini}, but the quality of fits to the susceptibility
(Fig~\ref{fig:chi1}), the non-linear susceptibility (Fig~\ref{fig:chi3},
left), and the metamagnetic transition  (Fig~\ref{fig:chi3}, right) is not
inferior to what was achieved earlier \cite{entropy}. 
However, our basic argument in favour
of the present scheme is not that it recovers standard results, but that it
allows the derivation of the phase diagram shown in Fig~\ref{fig:phase}. We
use the mean field decoupled hamiltonian
\begin{eqnarray}
{\cal H}_{\rm MF}  =  \Delta_1 |t_4\rangle \langle t_4| + 
\Delta_2 |t_2\rangle \langle t_2| +
\Delta_3 \sum_{\alpha =+,-}|d_{\alpha}\rangle \langle d_{\alpha}|
 \nonumber\\
 -g{\mu}_{\rm B}H J_z 
+ \lambda_{\rm oct}\left<{\cal T}_{z}^{\beta}\right>{\cal T}_{z}^{\beta}  
-\lambda_{\rm quad}\left<{\cal O}_{zx}\right>{\cal O}_{zx}
\label{eq:mean}
\end{eqnarray}
where $g=4/5$, and the octupolar mean field coupling constant 
$\lambda_{\rm oct}$ is meant to include the effective coordination number;
similarly for the quadrupolar coupling constant $\lambda_{\rm quad}$. 
We assume alternating octupolar order \cite{why}, and 
uniform ${\cal O}_{zx}$ order; 
the result would be the same if the high-field quadrupolar order 
is also alternating. We do not
introduce ${\cal O}_2^2$ or $\{J_x$, $J_y\}$ couplings, nevertheless 
$\langle {\cal O}_2^2\rangle \ne 0$ in the $B_1$ phase, and 
$\langle J_x\rangle\ne 0$ in the $E$ phase.  

At $H=0$, the only non-vanishing octupolar matrix element is 
$C=\langle t_1|{\cal T}_{z}^{\beta}|t_4\rangle \approx 8.8$.  Octupolar 
order is driven by  the large $C$: assuming $\lambda_{\rm oct}=0.336$K
we get the critical temperature $T_{O}(H=0)=17.2$K for 
${\cal T}_{z}^{\beta}$-type antiferro-octupolar order. 
Using a similar estimate, we find  
$\lambda_{\rm oc}^{\rm Np}\approx 0.2$K for NpO$_2$ which orders at 25K 
\cite{npo2}, thus the assumed  octupolar coupling strength is not 
unreasonable \cite{afm}.

The octupolar transition shows up as a break in the temperature 
derivative $(\partial \chi_1/\partial T)$ of the linear susceptibility
(Fig.~\ref{fig:chi1}, right). The sign of the discontinuity of slope is
related to the fact that the critical temperature decreases in magnetic field 
like $T_O(H)\approx T_O(H=0)-a_H H^2$ \cite{kf,sakak}. 
An Ehrenfest relation \cite{shah,kf} connects  the discontinuity of 
$(\partial \chi_1/\partial T)$  to that of the non-linear susceptibility 
$\chi_3$ (Fig.~\ref{fig:chi3}, left).

Up to the vicinity of the $t_1$--$d_-$ level crossing shown in
Fig.~\ref{fig:levels}, field effects can be understood within the 
$|t_1\rangle$--$|t_4\rangle$--$|t_2\rangle$ subspace. 
Applying ${\bf H} \parallel {\hat z}$ mixes $|t_2\rangle$ to  
$|t_1\rangle$. This has two effects. First, since ${\cal O}_2^2$ connects 
$|t_2\rangle$ to $|t_4\rangle$, the order parameter acquires a mixed 
${\cal T}_z^{\beta}$--${\cal O}_2^2$ character (cf. Table~\ref{tab:tab2}). 
Second, it reduces the octupolar matrix element, and thereby also $T_O$. 
With the parameters given before, the octupolar transition is fully suppressed
at $H_{{\rm cr},1}=34.7$T (see Fig.~\ref{fig:phase}). There is an
accompanying change in the slope of the magnetization curve which, however,
is not noticeable on the scale of Fig.~\ref{fig:chi3} (right).   

\begin{figure}[ht]
\centering
\includegraphics[height=5cm,angle=0]{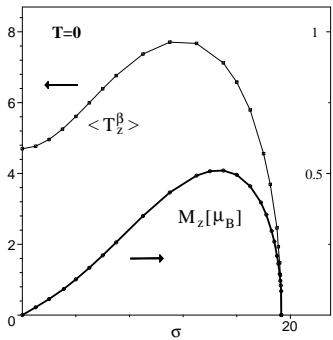}
\caption{Stress-induced magnetic moment in the octupolar phase. 
Thick line: $\langle M_z\rangle$ staggered magnetization, thin line: 
$\langle {\cal T}_z^{\beta}\rangle$ octupolar moment, as a function of the 
uniaxial pressure $\sigma\parallel [100]$ ($\sigma$ in arbitrary units).}
\label{fig:stress}
\end{figure}

A basis-independent description of field effects relies on the Landau
expansion of the Helmholtz potential ${\cal A}$ which yields the
magnetic field as a derived quantity ${\bf H} = (\partial {\cal A}/
\partial {\bf J})$ \cite{kf}. ${\cal A}$ is a sum of invariants. Two terms   
which are important for the present purpose, are  contained in
\begin{eqnarray}
{\cal I}(A_{2u}{\otimes}B_{1g}{\otimes}B_{2u}) & = & c_1
J_z({\bf 0}) {\cal T}^{\beta}_z ({\bf Q}){\cal O}_2^2(-{\bf Q})
\nonumber\\ 
& & +c_2 J_z({\bf Q}) {\cal T}^{\beta}_z (-{\bf Q}){\cal O}_2^2({\bf 0})\, .
\label{eq:inv}
\end{eqnarray}
$c_1$ and $c_2$ are non-zero even if we consider the lowest two levels only. 
The existence of this invariant can be exploited in several ways. 
In a uniform magnetic field, 
alternating octupolar order ${\cal T}^{\beta}_z$ 
induces similarly alternating quadrupolar order ${\cal O}_2^2$. 
Alternatively,  it follows that in the presence of uniform 
quadrupolar polarization  ${\cal O}_2^2$, alternating octupolar order gives 
rise to a magnetic moment $J_z$ with the same periodicity. 
Such a quadrupolar polarization is created by 
uniaxial stress applied in the [100] 
direction, which is observed to give rise to alternating
magnetic moments of $O(0.1\mu_{\rm B})$, clearly different from 
micromagnetism \cite{yokoyama}. Fig.~\ref{fig:stress} shows the stress-induced
staggered magnetization  for the same set of parameters as in 
previous plots. Sufficiently large stress suppresses octupolar order, 
like a sufficiently strong field does. The maximum induced moment is 
$\sim 0.5\mu_{\rm B}$; the measured $\sim 0.2\mu_{\rm B}$ \cite{yokoyama} 
may belong to the rising part of the curve.

Stress applied along the $z$-axis induces ${\cal O}_2^0$
which transforms according to 
the identity representation $A_{1g}$, thus it does not appear in the 
invariants, and it is not predicted to induce magnetism.  

We note that the $A_{1u}$ triakontadipole 
${\overline{J_x J_y J_z (J_x^2 - J_y^2)}}$ (see Table~\ref{tab:tab1}) 
would not give rise to
stress-induced magnetism and is therefore not a suitable choice 
as order parameter in the limit $H\to 0$. 

We now discuss the high-field behavior at $H>H_{{\rm cr},1}$. The single-ion 
levels $t_1$ and $d_-$ would cross at $H_{\rm cross}=37.3$T. 
Since $|t_1\rangle$ and $|d_-\rangle$ are connected by $E$ operators  
including ${\cal O}_{zx}$, a range of fields centered on  $H_{\rm cross}$
is certain to favour $\{ {\cal O}_{zx},{\cal O}_{yz} \}$ quadrupolar order,  and 
simultaneous $\{ J_x, J_y \}$ dipolar order. We chose a weak quadrupolar 
interaction $\lambda_{\rm quad}=0.054$K in Eqn.~(\ref{eq:mean}); this
gives quadrupolar order between the critical fields $H_{{\rm cr},2}=35.8$T and 
$H_{{\rm cr},3}=38.8$T. The amplitude of quadrupolar order is not small
(Fig.~\ref{fig:phase}) but the  ordering temperature is low ($\sim 1$K) because the coupling 
is weak. The $E$ phase shows up as the steep part of the magnetization curve in 
Fig.~\ref{fig:chi3} (right). For $\lambda_{\rm quad}=0$ we would have a jump-like metamagnetic transition at  $H=H_{\rm cross}$.

In summary, our crystal field scheme gives an $H$--$T$ phase diagram in overall agreement  
with experiments. Time reversal 
invariance breaking by the ${\cal T}_z^{\beta}$ 
octupolar order in the low-field phase is essential to allow the prediction of large-amplitude 
antiferromagnetism induced by transverse uniaxial stress. The disjoint high-field phase has mixed 
quadrupolar--dipolar character.

\begin{acknowledgements}
The authors are greatly indebted to Yoshio Kuramoto for a most  enlightening discussion and continuing encouragement. We were supported by the Hungarian national grants OTKA T 038162, and OTKA T 037451.
\end{acknowledgements}


\begin{thebibliography}{99}
\bibitem{xyz} A model based on  ${\cal T}_{xyz}$ ordering would give
rather similar results.
\bibitem{entropy} Since our scheme works with five states, it may seem to
  overestimate the entropy \cite{schlabitz}. However, magnetic field and
intersite interactions  split off the lowermost two or three states,
thus we have effectively a three-state model.
\bibitem{why} The symmetry happens to coincide with that of the micromagnetic order which 
 we ascribe to a minority phase. Our choice of alternating 
order corresponds to that of stress-induced 
 strong antiferromagnetism \cite{yokoyama}.
\end{thebibliography}
\end{document}